\documentclass[11pt]{article} 

\usepackage[utf8]{inputenc} 

\usepackage{geometry} 
\usepackage{graphicx} 

\usepackage{authblk}
\usepackage{booktabs} 
\usepackage{array} 
\usepackage{paralist} 
\usepackage{verbatim} 
\usepackage{subfig} 
\usepackage[square,comma,sort&compress]{natbib}%
\usepackage[toc,section=section]{glossaries}%
\usepackage[font=small,labelfont=bf]{caption}
\usepackage{xspace}
\usepackage{hyperref}%
\usepackage{paralist}%
\usepackage{subfig}
\usepackage{url}%

\usepackage{fancyhdr} 
\usepackage{sectsty}
\allsectionsfont{\sffamily\mdseries\upshape} 

\usepackage[nottoc,notlof,notlot]{tocbibind} 
\usepackage[titles,subfigure]{tocloft} 

\newacronym{bubsim}{BuBSim}{Bartz und Bartz Simulator}
\newacronym{des}{DES}{Discrete Event Simulation}
\newacronym{doe}{DOE}{Design of Experiments}
\newacronym{icu}{ICU}{Intensive Care Unit}
\newacronym{rmse}{RMSE}{Root Mean Square Error}
\newacronym{smbo}{SMBO}{Surrogate-Model Based Optimization}
\newacronym{spo}{SPO}{Sequential Parameter Optimization}
\newacronym{spot}{SPOT}{Sequential Parameter Optimization Toolbox}
\glstoctrue
\makeglossaries

\newcommand{\bubsim}{\gls{bubsim}\xspace}
\newcommand{\des}{\gls{des}}
\newcommand{\doe}{\gls{doe}}
\newcommand{\icu}{\gls{icu}}
\newcommand{\rmse}{\gls{rmse}\xspace}
\newcommand{\smbo}{\gls{smbo}\xspace}
\newcommand{\spot}{\gls{spot}\xspace}

\renewcommand{\cite}{\citep}

\title{Optimization of High-dimensional Simulation Models Using Synthetic Data}
\author[1]{Thomas~Bartz-Beielstein}
\author[2]{Eva Bartz} 
\author[1]{Frederik Rehbach} 
\author[2]{Olaf Mersmann}

\affil[1]{Institute for Data Science, Engineering, and Analytics, TH Köln, Germany}
\affil[2]{Bartz \& Bartz GmbH, Gummersbach, Germany}

\affil[ ]{\textit {\href{mailto:thomas.bartz-beielstein@th-koeln.de}thomas.bartz-beielstein@th-koeln.de}}

\begin{document}
\maketitle

\begin{abstract}
 \noindent Simulation models are valuable tools for resource usage estimation and capacity planning.
 In many situations, reliable data is not available. We introduce the BuB simulator, which
 requires only the specification of plausible intervals for the simulation parameters. By performing a surrogate-model based optimization,
 improved simulation model parameters can be determined. Furthermore, a detailed statistical analysis can be performed, which 
 allows deep insights into the most important model parameters and their interactions.
 This information can be used to screen the parameters that should be further investigated.
 To exemplify our approach, a capacity and resource planning task for a hospital was simulated and optimized. 
 The study explicitly covers difficulties caused by the COVID-19 pandemic. It can be shown, that even if only limited real-world data is available,
 the BuB simulator can be beneficially used to consider worst- and best-case scenarios. 
 The BuB simulator can be extended in many ways, e.g., by adding further resources (personal protection equipment, staff, pharmaceuticals) or by specifying several cohorts (based on age, health status, etc.).
 
 \end{abstract}

\section{Introduction}
Simulation models are valuable tools for resource usage estimation and capacity planning.
They can either be implemented top down, e.g., using time-series approaches~\cite{Hynd08b} or bottom-up, e.g., using discrete-event simulation~\cite{Bank01a}.
We present a bottom-up approach: the \bubsim is a discrete-event simulation model, which is applied to a hospital resource planning problem.
The project is motivated by the current COVID-19 pandemic. Health departments can use \bubsim to forecast demand for \icu\/ beds, ventilators, and  staff resources.

Health departments are facing a very demanding situation, because the development of the COVID-19 pandemic is unknown. 
No experiences about the required resources exist.
Health systems in some countries  collapsed, whereas in other countries, no severe resource problems could be observed.
Resource planning is of great importance in these situations.
Unfortunately, if there is a new outbreak, no reliable data ist available.

To tackle the problem of missing data, we present an approach that relies on \emph{synthetic data}. It combines optimization and simulation techniques to build an adaptive model.
Our approach combines two powerful  technologies: (a) \des\/ and (b) \smbo~\cite{Jin01a}.
By combining these, we are able to build a simulation model that only requires \emph{plausible intervals} for the model parameters.
\bubsim has more than 30 internal model parameters which are optimized during the simulations.

The paper is structured as follows: 
Section~\ref{sec:bubsim} briefly introduces the \bubsim model.
After introducing \bubsim, strategies for the optimization of high-dimensional simulation models will be discussed.
Section~\ref{sec:experiments} describes the experiments, i.e., simulation data, optimization criteria, parameters to be optimized, and the algorithms.
Section~\ref{sec:discussion} discusses the results with a special focus on optimized parameters and sensitivity analysis.
The paper concludes with an outlook in Section~\ref{sec:outlook}, which discusses next steps.

\section{The \bubsim\/ Model}\label{sec:bubsim}
\citet{Law07a} considers three distinctions between simulation approaches: (a) deterministic or stochastic, (b) static or dynamic, and (c) continuous or discrete. 
\des\/  is an approach for modeling stochastic, dynamic and discretely changing systems. 

The \emph{state}\/ of such systems is defined to be a collection of variables necessary to describe the system at any time~\cite{Bank01a}.
The term \emph{discrete}\/ refers to the characteristic behavior of its components, which can be described as \emph{events}: an event is an instantaneous
 occurrence that may change the state of the system~\cite{Bank01a}. Between events, all the state variables remain constant.
Benefits of \des\/ are manifold and range from 
providing insights into the process’ risk and efficiency to
estimating the effects of alternating configurations of the system.
It helps to gain insight into consequences of redesign strategies.

\des\/ has been successfully applied to problems that model  
customers arriving at a bank,   
products being manipulated in a supply chain, 
and
the performance of configurations of a telecommunications network~\cite{Bank01a}.

\begin{figure}
\centering
\includegraphics[width=0.5\linewidth]{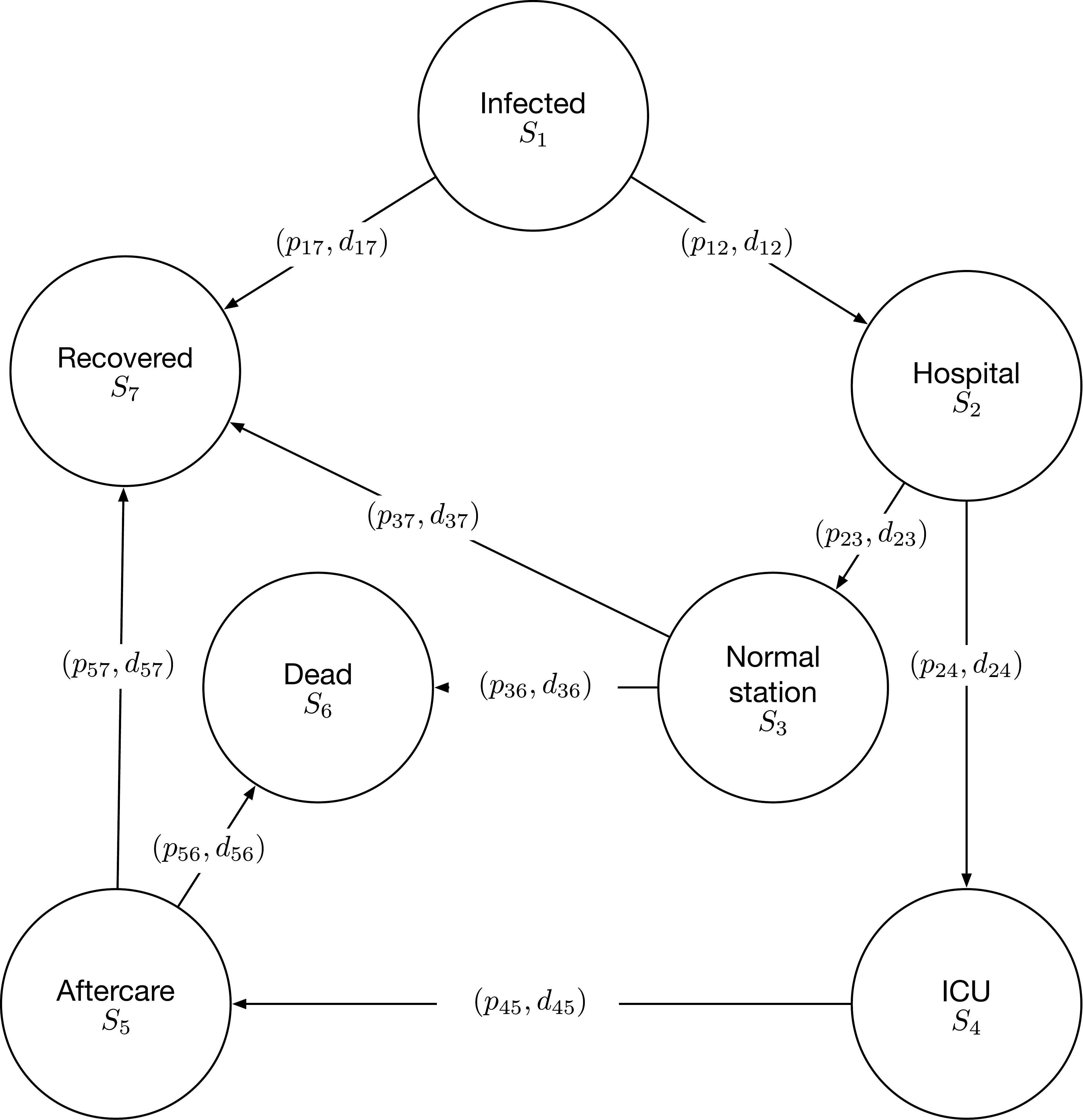}
\caption{Simplified model of patient flows in a hospital. Nodes represent states ($S_i$). Edges represent state changes with associated probabilities ($p_{ij}$) and durations ($d_{ij})$.
Probabilities and durations at time step $t$ of the optimization will be referred to as \emph{model parameters}\/ $\vec{x}_t$.}
\label{fig:hospital}
\end{figure}

\bubsim\/ models patient flows in hospitals and focusses specifically on COVID-19 patients.
Each patient follows a path, i.e., after a state-specific duration $d_{ij}$, he moves from state $S_i$ with a certain probability $p_{ij}$ to the next state $S_j$.
A graph can be used to model this behavior (Fig.~\ref{fig:hospital}). 
For example, an infected patient (state $S_1$) goes after $d_{12}$ days with probability $p_{12}$ to the hospital (state $S_2$). With probability $p_{17}$, he recovers (state $S_7$) after $d_{17}$ days without visiting a hospital. The probabilities of outgoing nodes sum to 1, e.g., $p_{17} = 1 - p_{12}$.
The \bubsim\/ model used in our study requires the specification of more than 30 parameters.

\section{Simulations}\label{sec:experiments}

\subsection{Simulating Infections}
Arrival times, as in queuing systems, were used to model the occurrences of COVID-19 infections. A Poisson distributed random variable was used to generate these events.
Node $S_1$ represents the state \emph{infected}, which is the starting point of our patient-flow simulations.
The Poisson distribution was selected, because it is a discrete probability distribution that expresses the probability of a given number of events occurring in a fixed interval of time. 
Furthermore, the Poisson distribution is applied to simulations with a large number of relatively rare events.
For the simulation of the infection process, we assume that the infection events occur independently of the time since the last infection with a constant mean $\lambda$.
We considered the following setting: every day, on average, four new infections occur. The simulation period ranges from the 1st of September 2020 until the 30th of November 2020, which results in a period of $T =  91$ days. In addition to the ``base'' infection rate, we considered \emph{peak events}, i.e., a larger number of infections that occur during a short time period, e.g., one day.
The variable $u_t$ is used to model the infections at day $t$.
The scenario is illustrated in the left panel of Fig.~\ref{fig:beds}.

\subsection{Model Parameters: $\vec{x}_t$}
Plausible values from the literature were used as starting points for the durations $d_{ij}$ and probabilities $p_{ij}$ to model the state changes.
The \bubsim\/ modeling framework does not need exact values. The specification of \emph{plausible intervals}\/ is recommended, because \bubsim\/ tries to optimize the values of the model parameters
($p_{ij}, d_{ij})$. Therefore, we specified upper and lower bounds for every model parameter.
 The vector of model parameters at each time step will be denoted as $\vec{x}_t$. The index $t$ reflects the time dependency of this parameter.

\subsection{Ground Truth, Resources: $R_i$}
\begin{figure}
\centering
\includegraphics[width=0.45\linewidth]{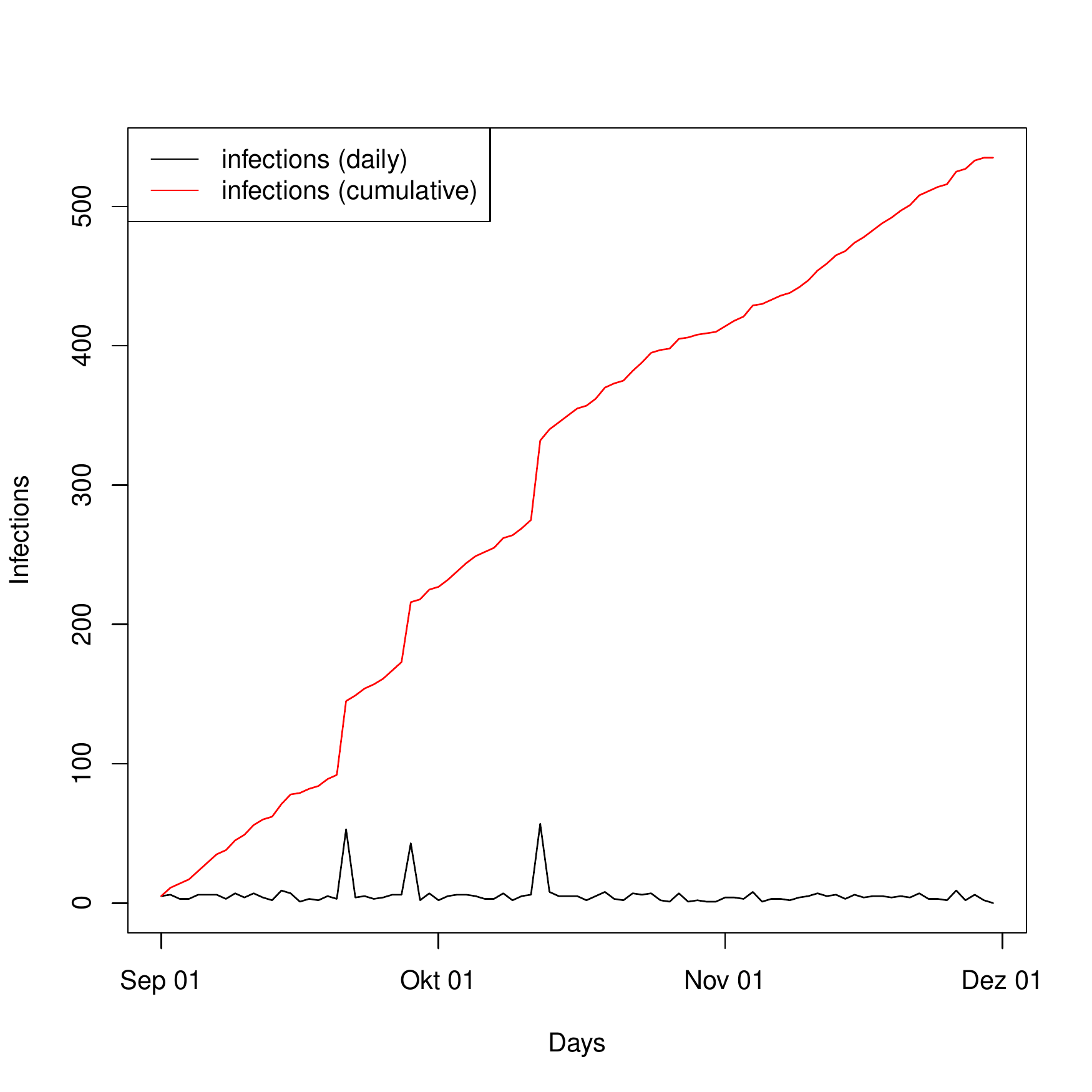}
\hskip1em
\includegraphics[width=0.45\linewidth]{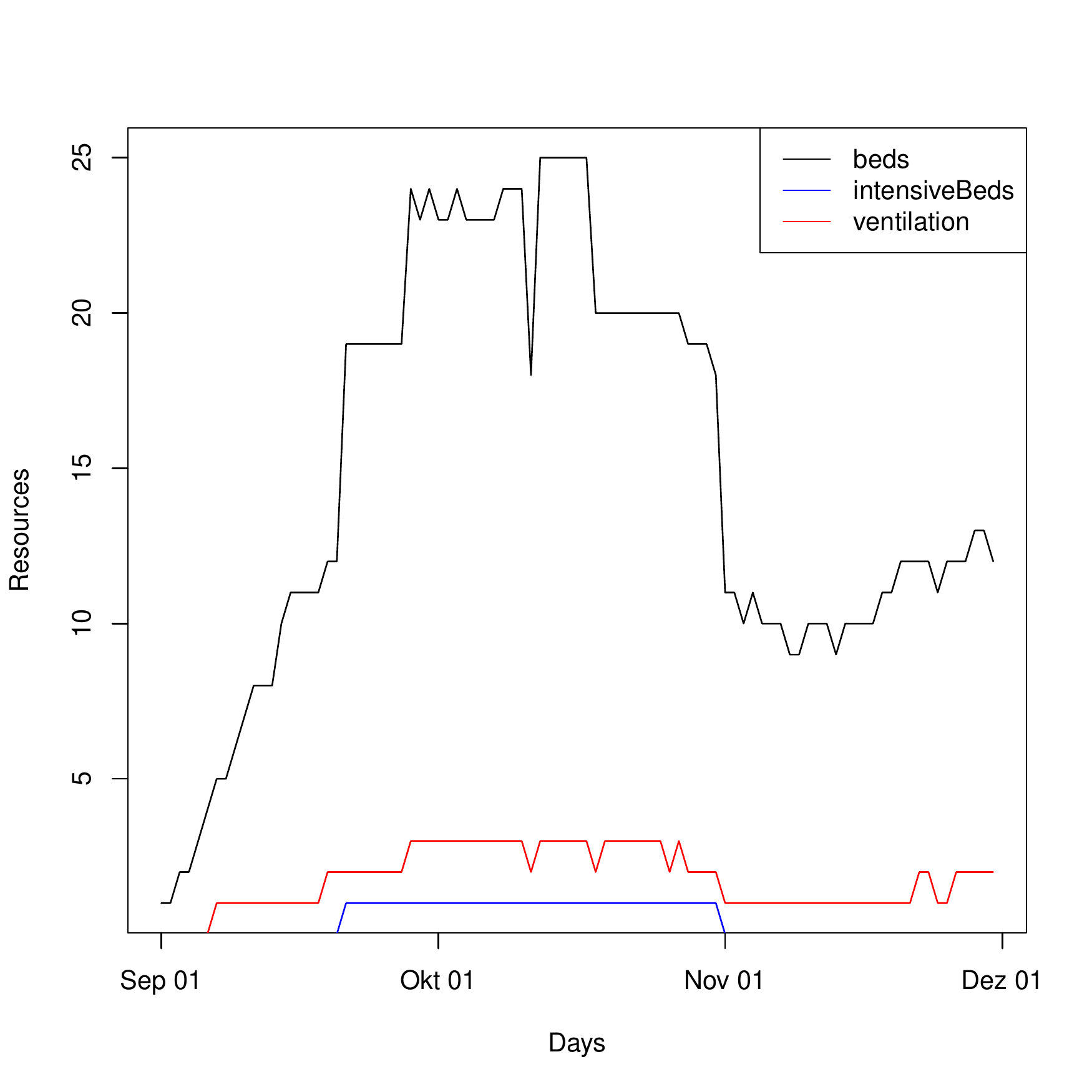}
\caption{\emph{Left:}\/ Infections. Synthetic data. Realization of Poisson-distributed random variables with three peak events. 
\emph{Right:}\/ Resource usage: beds, \icu\/ beds,  and \icu\/ beds with ventilators, i.e., $R_{\text{bed}}$,   $R_{\text{icu}}$, and $R_{\text{vent}}$, respectively. Synthetic data. These values should be estimated by the simulation model, see Equation~\ref{eq:haty}.}
\label{fig:beds}
\end{figure}

The \bubsim model, as used in our study, models three resources:
\begin{itemize}
  \item regular hospital beds (denoted bed)
  \item \icu~beds without ventilation (denoted icu)
  \item \icu~beds with ventilation (denoted vent)
\end{itemize}
Because no real-world data is available, we estimated the ``ground truth'' , i.e., the observed resource usage, as follows:
\begin{enumerate}
\item Using the synthetic data $\{u_t\}$, $t = 1, \ldots, T$,  we calculated the number of infected individuals during a time window of two weeks. Let $U_{14}(t)$ denote this number for day $t$:
\begin{equation}
U_{14}(t) = \sum_{i=0}^{14} u_{t - i}.
\end{equation}
This is also possible if real data are available, because many institutions, e.g., Johns~Hopkins~University\footnote{\url{https://coronavirus.jhu.edu}}, publish their statistics on a daily basis. 
\item The number of individuals, who are hospitalized and require a certain type of bed is calculated as a percentage of the infected individuals. Let $r_{\text{bed}}$,   $r_{\text{icu}}$, and $r_{\text{vent}}$ denote the percentage of regular hospital beds,  \icu\/  beds without ventilation, and \icu\/ beds with ventilation, respectively. These depend heavily on the local situation. They can roughly be estimated from figures published by regional health organizations.
Then we can calculate the ``ground truth'' for the resource allocations at time step $t$ as follows:
\begin{equation}\label{eq:ground}
R_i(t) = r_i \times U_{14}(t),
\end{equation}
where $i \in 
\{
\text{bed},  \text{icu}, \text{vent}
\}$.
The right panel in Fig.~\ref{fig:beds} illustrates the resource allocations over time.
\end{enumerate}

\subsection{Simulations}
As illustrated in Fig.~\ref{fig:optimization}, the \bubsim simulator requires only two input parameters:
\begin{enumerate}
\item $\vec{x}_t$, the model parameters
\item $\vec{u}_t$, the number of infections.
\end{enumerate}
Based on these two inputs, \bubsim estimates the required resources---in our case, the beds, \icu\/ beds,  and \icu\/ beds with ventilators, i.e., $\hat{R}_{\text{bed}}$,   $\hat{R}_{\text{icu}}$, and $\hat{R}_{\text{vent}}$, respectively.
The simulation output, i.e, the required resources on each day $t$ will be denoted as $\hat{\vec{y}}_t$, i.e., 
\begin{equation}\label{eq:haty}
\hat{y}_t = \left( \hat{R}_{\text{bed}}(t),   \hat{R}_{\text{icu}}(t), \hat{R}_{\text{vent}}(t) \right)
\end{equation}

\begin{figure}
\centering
\includegraphics[width=0.45\linewidth]{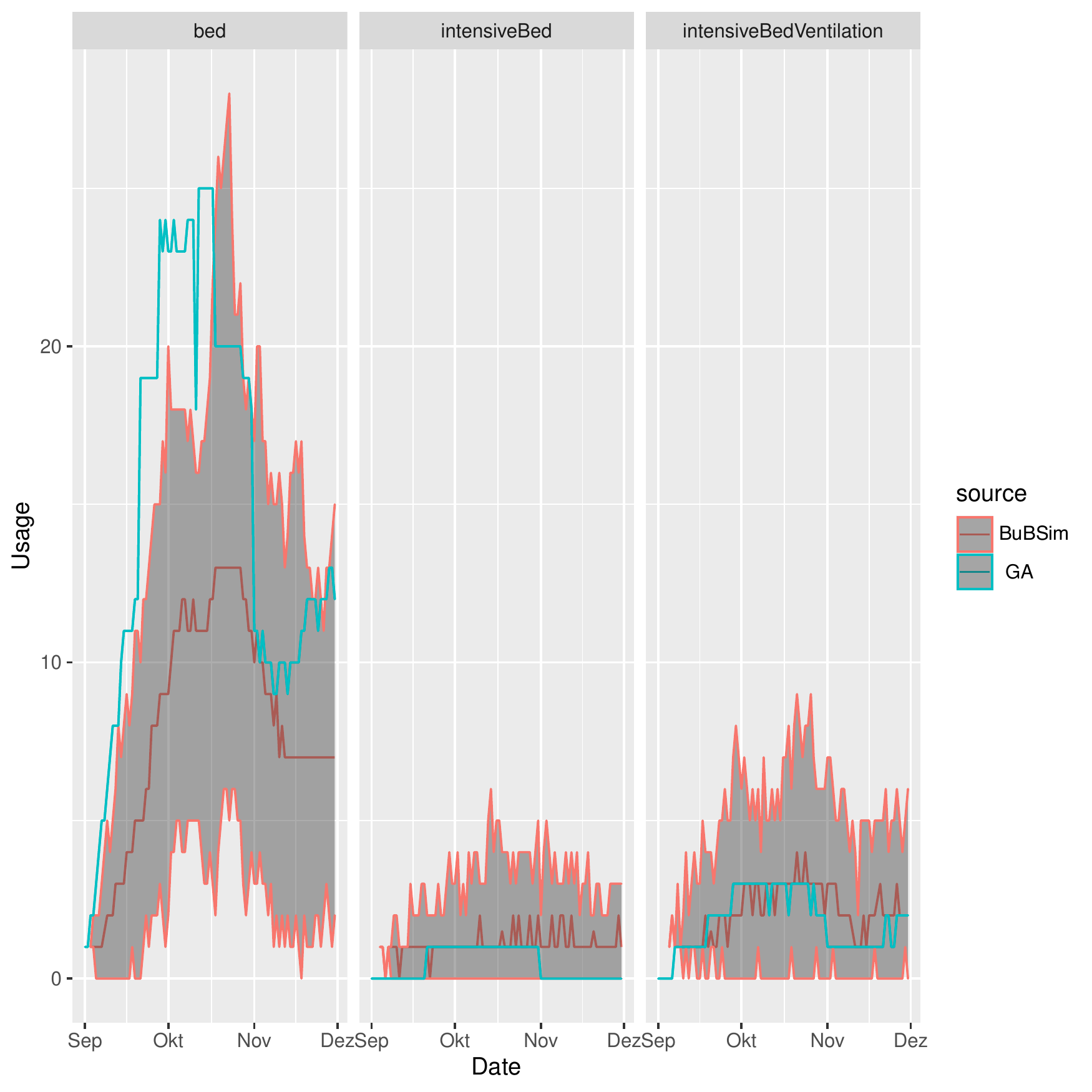}
\hskip1em
\includegraphics[width=0.45\linewidth]{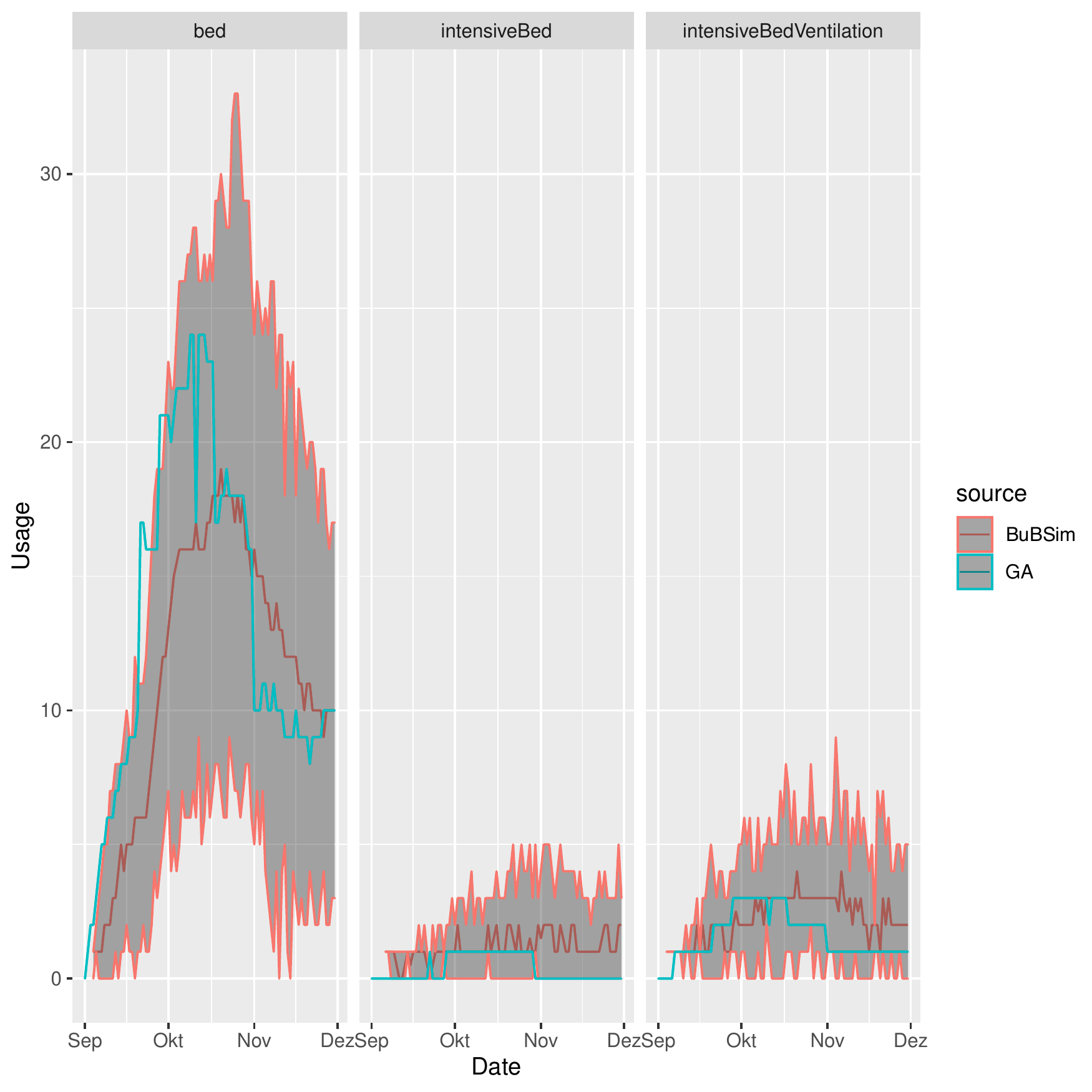}
\caption{ \emph{Left:} Simulation of the demand using default model parameters. 
The three subplots show the demand for each resource, i.e., for each type of bed ($\hat{R}_{\text{bed}}$,   $\hat{R}_{\text{icu}}$, and $\hat{R}_{\text{vent}}$).
 \emph{Dark red} lines represent the median from $n$ simulation runs. 
  \emph{Turquoise} lines show the ground truth, i.e., the observed values ($R_{\text{bed}}$,   $R_{\text{icu}}$, and $R_{\text{vent}}$) that were calculated with Equation~\ref{eq:ground}. 
 \emph{Right:} Same setting, if the simulation is performed with the optimized parameters. 
  Optimization reduces the \rmse from 9.16 to 6.81.}
\label{fig:default}
\end{figure}

Results from the simulation with default model parameters $\vec{x}_t$ are shown in Fig.~\ref{fig:default}.

\subsection{Optimization}
\subsubsection{Optimization Goals}

\begin{figure}
\centering
\includegraphics[width=0.7\linewidth]{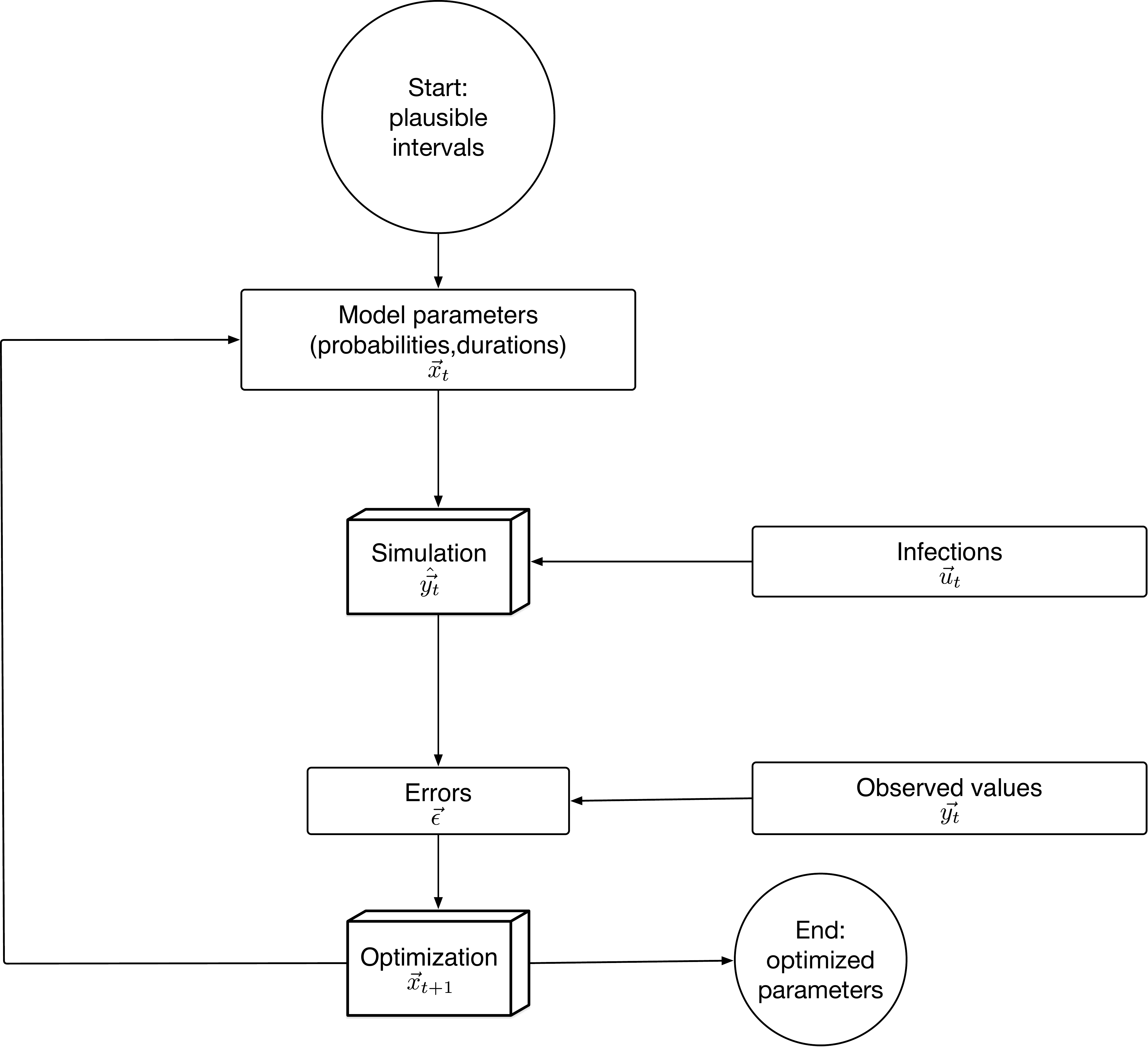}
\caption{Simulation via optimization.}
\label{fig:optimization}
\end{figure}

Based on the simulation data, optimization runs can be performed to estimate reasonable values. The optimization via simulation loop is shown in Fig.~\ref{fig:optimization}.
The \rmse as shown in Eq.~\ref{eq:rmse}, is used to measure the error of the simulator.
We formulate a \emph{minimization} problem, because smaller errors, $\epsilon$, are better:
\begin{equation}\label{eq:rmse}
  \epsilon 
  = 
  \sum_{k\in\{\text{bed},\text{icu},\text{vent}\}}
    w_k  \sqrt{\frac{1}{T} \sum_{t=1}^T \left(R_k(t) - \hat{R}_k(t)\right)^2}
\end{equation}
Here, $T$ denotes the duration of the simulation, i.e., the number of days simulated and $k$ the three different categories of beds required.
A weighted approach to reflect different importance of the different bed categories can easily be implemented.
In our experiments, $w_k = 1/3$.

\subsubsection{Sequential Parameter Optimization}
The \spot provides sophisticated statistical and machine learning tools for \doe, optimization, and sensitivity analysis~\cite{BLP05}. 
It is based on \smbo~\cite{Bart16n}.
\begin{table}
\caption{Regression analysis. Shown is the subset of the most important model parameters resulting from a step-wise regression analysis.}
\label{tab:reg}
\centering
\begin{tabular}{rrrrrr}
  \hline
Variable & Meaning  & Estimate & Std. Error & t value & Pr($>$$|$t$|$) \\ 
  \hline
& (Intercept) & 11.5544 & 0.4027 & 28.70 & 0.0000 \\ 
  $x_1$  & DaysInfectedToHospital & 0.1767 & 0.0138 & 12.78 & 0.0000 \\ 
  $x_2$ & DaysNormalToHealthy & -0.1723 & 0.0101 & -17.08 & 0.0000 \\ 
  $x_{16}$ & GammaShapeParameter & -2.1465 & 0.1330 & -16.13 & 0.0000 \\ 
  $x_{19}$ & PercentageHospitalToVentilation & 13.1687 & 2.9242 & 4.50 & 0.0000 \\ 
 \hline
\end{tabular}
\end{table}

\subsubsection{Optimization Results}
To illustrate the impact of the \spot optimization, we performed two simulations of the patient flow. 
The first simulation used the default model parameters, which are based on the \emph{plausible intervals}.
The second simulation uses the optimized parameter set that was obtained with \spot. 
Optimization reduces the \rmse  by more than 25 percent (from 9.16 to 6.81). The result is illustrated in the right panel of Fig.~\ref{fig:default}

\subsubsection{Sensitivity Analysis}
The regression analysis is shown in Table~\ref{tab:reg}. Results from the regression analysis can be interpreted as follows:
to reduce the model error, the amount of days, i.e., $x_1$, until a patient goes to the hospital, should be reduced. 
And, the estimated number of days $(x_2)$ until a patient leaves the normal station and  recovers, should be increased.
Furthermore, the value of the  shape parameter of the Gamma distribution, which is used to model the durations $(d_{ij})$, should be increased.
Finally, $x_{19}$, the percentage of patients that  need an \icu\/ bed with ventilation when they arrive at the hospital, should be increased.

\begin{figure}
\centering
\includegraphics[width=0.7\linewidth]{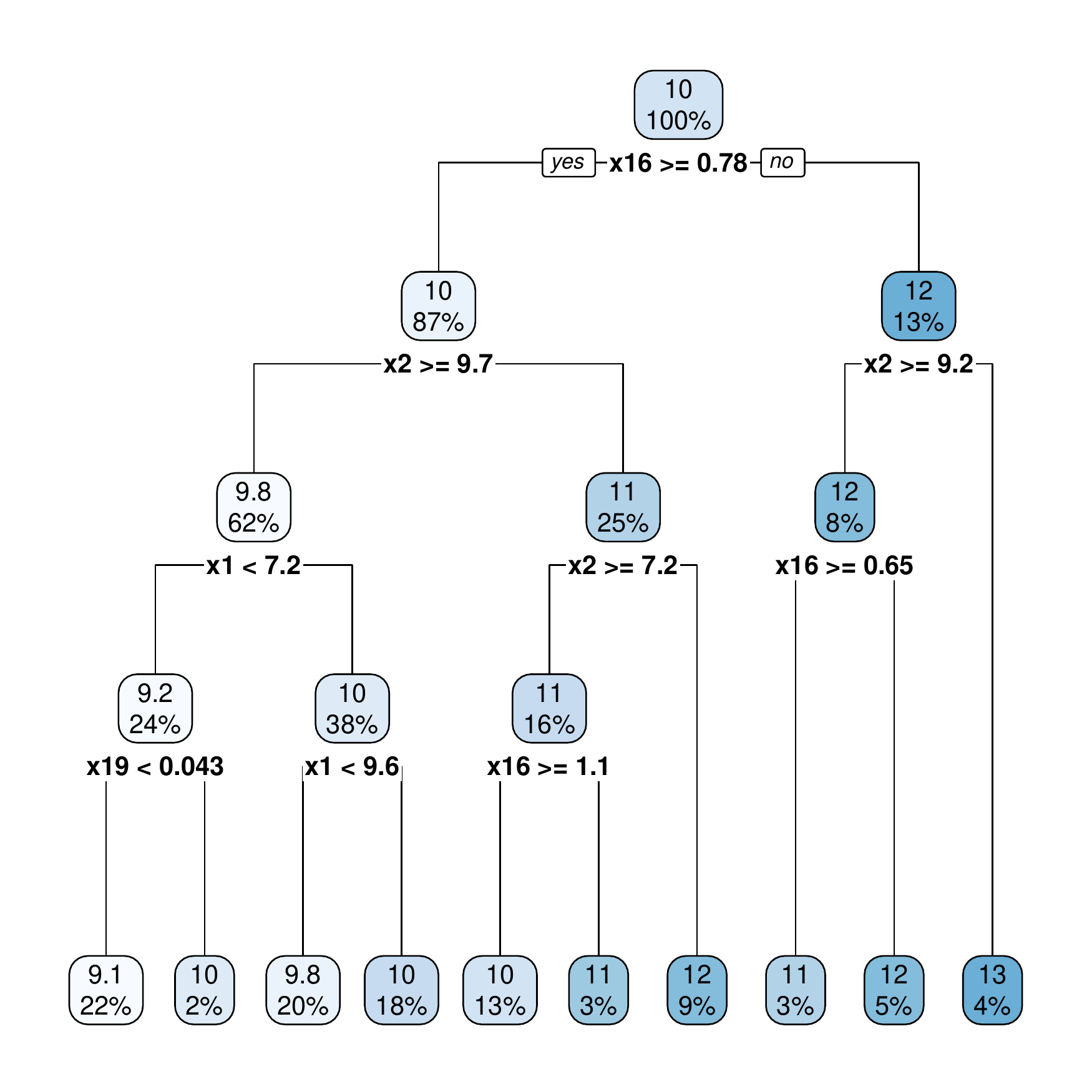}
\caption{Regression tree analysis. The tree-based model reveals, that the same model parameters as in the regression model, are of importance: $x_1$, 
  $x_2$,  $x_{16}$, and $x_{19}$.}
\label{fig:tree1000}
\end{figure}

In addition to the regression analysis, a tree-based analysis can be performed. The corresponding regression tree is shown in Fig.~\ref{fig:tree1000}.
The tree-based analysis confirms the results from the regression analysis, because the same variables are considered important: $x_1$, 
  $x_2$,  $x_{16}$, and $x_{19}$.
\begin{figure}
\centering
\includegraphics[width=0.5\linewidth]{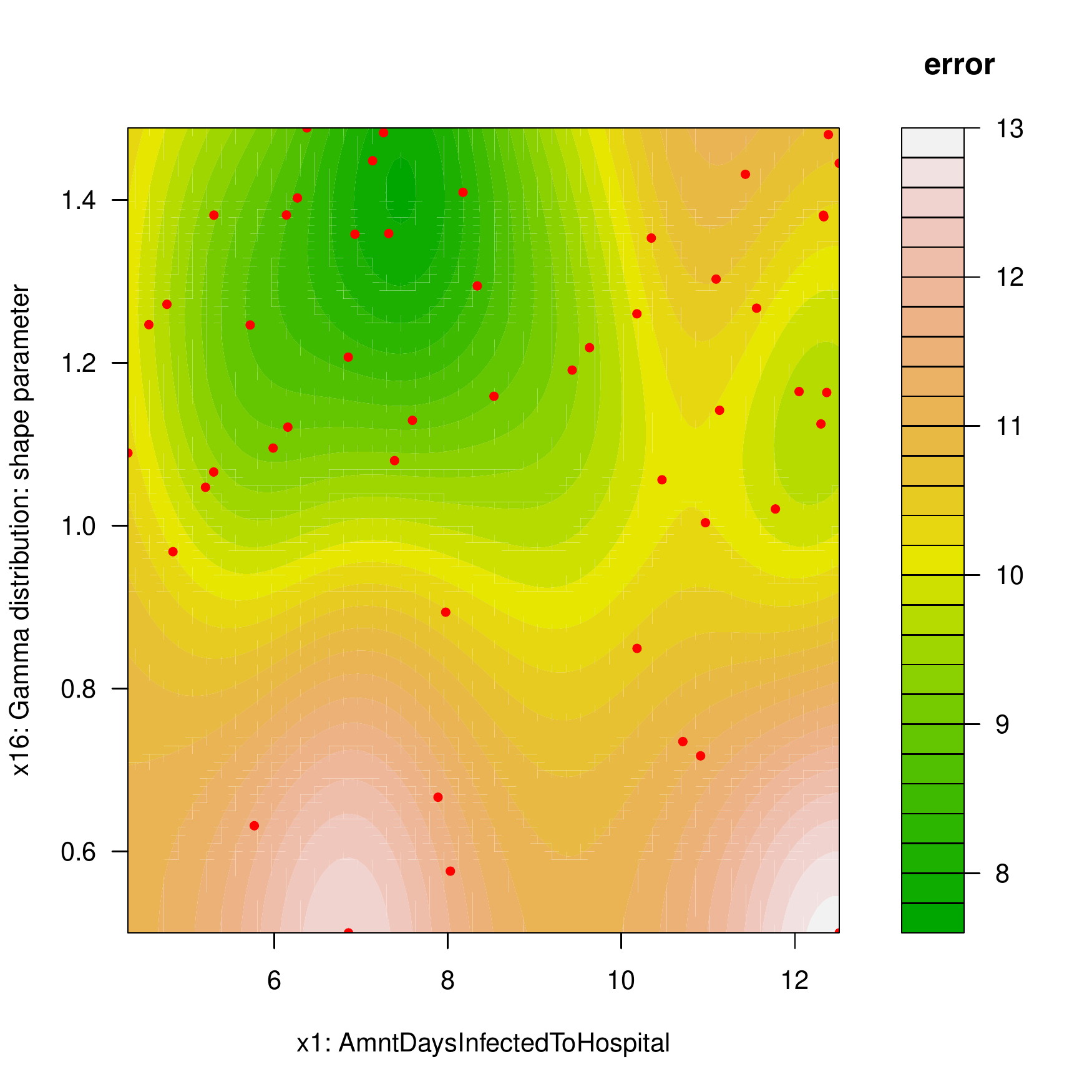}
\caption{Contour plot. Values of the error $\epsilon$ (introduced in Equation~\ref{eq:rmse}) as a function of the shape parameter of the  Gamma distribution ($x_{16}$) and the average number of days after which infected patients arrive at the hospital ($x_1$).}
\label{fig:landscape1000}
\end{figure}
Furthermore, \spot offers visualization tools and  statistical procedures for investigating variable importance, interactions, and many aditional functions~\cite{Bart17parxiv}.
Figure~\ref{fig:landscape1000} plots the relationship between the shape parameter of the Gamma distribution ($x_{16}$) and the number of days it takes infected patients to arrive at the hospital ($x_1$).
This figure supports the results from the regression and tree-based analysis: $x_{16}$ should be increased while $x_1$ should be reduced.

\section{Discussion}\label{sec:discussion}

\subsection{The \bubsim Simulator}
We introduced a novel conceptual framework that allows reliable and detailed resource planning for hospitals.
To exemplify our approach, we focused on simulating bed capacities.
The \bubsim model that was implemented in our study modeled three different bed categories:
\begin{itemize}
\item hospital beds
\item \icu\/ beds without ventilation
\item \icu\/ beds with ventilation.
\end{itemize}

\subsection{Synthetic Data}
Most importantly, the \bubsim model can be run without any real-world data (although it would benefit from it). 
\bubsim only requires the specification of \emph{plausible intervals} for the model parameters $\vec{x}_t$ and infection data $\vec{u}_t$.
Using the sophisticated optimization toolbox \spot, improved parameters are found.
\bubsim allows the simulation of different infection scenarios to analyze both \emph{best-} and \emph{worst-case} scenarios.
In particular, this allows practitioners to simulate the impact of \emph{peak} and \emph{superspreader events} on the bed capacities.

\subsection{Validation and Automated Learning}
If real-world data is available, e.g., from local hospitals, it could be used to further refine the model. 
\bubsim allows easy integration of additional data sources.
Furthermore, \spot and \bubsim provide interfaces to integrate \emph{dynamic} data (data streams). 
Data from online sources, e.g., Johns Hopkins, Robert Koch Institute, and others, can be fed directly into the \bubsim simulator. 
This enables dynamic and continuous updates of the model without any intervention.

\section{Outlook}\label{sec:outlook} 

\begin{description}
\item{Comparability:}
To demonstrate the usefulness of the \spot approach, a comparison with other optimization tools is of great interest. This comparison will be subject of a forthcoming study, which will use a portfolio that includes  optimization approaches based on simulated annealing, Nelder-Mead, quasi-Newton and conjugate-gradient algorithms.

\item{Validation:}
\bubsim and \spot can show their strengths, if real data is available. Anyhow, even if this is not available, the simulation environment can be fed with synthetic data and provide valuable insights into 
future developments (resource usage for certain scenarios).

\item{Extensibility:}
Further resources can easily be added. Until now, we considered three different bed categories. In addition, different resources like personal protection equipment, staff, pharmaceuticals, technical equipment (various ventilators),
hospital facilities, etc. can be modeled.
In addition, several cohorts (based on age, health status, etc.) can be implemented.

\item{Applicability:}
This simulator is specifically designed for hospitals, health departments, COVID-19 crisis teams, decision makers in politics and many others.
It is applicable to many different scenarios that require predictive planning.
\end{description}

\bibliographystyle{plainnat} 
\bibliography{bart20j}
\printglossary
\end{document}